\documentclass[aps,prl,twocolumn,floats,showpacs]{revtex4}
\usepackage{bbm,epsfig,citesort}

\newcommand{\1}{{\mathbbm{1}}}

\newcommand{\p}{\partial}
\newcommand{\e}{\vec e}
\begin{document} \draft 
\title{Magnon-mediated Binding between Holes in an Antiferromagnet}
\author{C.\ Br\"ugger, F.\ K\"ampfer, M.\ Pepe, and U.-J.\ Wiese}
\affiliation{Institute for Theoretical Physics, Bern University,
Sidlerstrasse 5, 3012 Bern, Switzerland}
\date{February 9, 2006}
%\twocolumn[\hsize\textwidth\columnwidth\hsize\csname @twocolumnfalse\endcsname

\begin{abstract}
The long-range forces between holes in an antiferromagnet are due to magnon 
exchange. The one-magnon exchange potential between two holes is proportional 
to $\cos(2 \varphi)/\vec r \, ^2$ where $\vec r$ is the distance vector of the 
holes and $\varphi$ is the angle between $\vec r$ and an axis of the square 
crystal lattice. One-magnon exchange leads to bound states of holes with 
antiparallel spins resembling $d$-wave symmetry. The role of these bound states
as potential candidates for the preformed Cooper pairs of high-temperature 
superconductivity is discussed qualitatively. 
\end{abstract} 

\pacs{12.39.Fe, 74.20.Mn, 75.30.Ds, 75.50.Ee}
\maketitle
Over the past twenty years, understanding the dynamical mechanism responsible 
for high-temperature superconductivity \cite{Bed86} has remained a great 
challenge in condensed matter physics. Unfortunately, microscopic 
systems such as the Hubbard or $t$-$J$ model, which may indeed contain the 
relevant physics, have thus far neither been solved analytically nor 
numerically beyond half-filling. While analytic solutions suffer from 
uncontrolled approximations, numerical simulations suffer from the fermion sign
problem. Although there have been numerous attempts to understand 
high-temperature superconductors via their undoped antiferromagnetic precursors
\cite{Miy86,Sca86,And87,Gro87,Tru88,Sch88,Wen89,Sha90,Mon91,Poi94}, the 
dynamical role of spin fluctuations remains a controversial issue. In 
particular, there seems to be no agreement if two holes doped into an 
antiferromagnet can form a bound state or not. In the following we will show
that --- by one-magnon exchange --- two holes can indeed form an infinite 
number of bound states. While it remains to be seen if this may lead to an 
explanation of high-temperature superconductivity, our results shed light on 
the mechanism responsible for the formation of charge pairs in the 
antiferromagnetic phase.

The low-energy physics of antiferromagnets is governed by their Goldstone 
bosons --- the magnons \cite{Cha89,Neu89,Fis89}. In analogy to chiral 
perturbation theory \cite{Gas85} --- the effective theory for the pions of the 
strong interactions \cite{Wei79} --- a systematic magnon effective theory was 
constructed in \cite{Has90,Has93}. Analogies between pion and magnon dynamics
have been investigated in \cite{Bae04}. The magnon field can be represented by 
a $CP(1)$ projection matrix
\begin{equation}
P(x) = \frac{1}{2}\left[\1 + \e(x) \cdot \vec \sigma\right],
\end{equation}
that obeys $P(x)^\dagger = P(x)$, $\mbox{Tr} P(x) = 1$, and $P(x)^2 = P(x)$. 
Here $x = (\vec x,t)$ is a point in $(2+1)$-d space-time,
\begin{equation}
\vec e(x) = 
(\sin\theta(x) \cos\varphi(x),\sin\theta(x) \sin\varphi(x),\cos\theta(x))
\end{equation}
is the staggered magnetization, and $\vec \sigma$ are the Pauli matrices. Under
global spin rotations $g \in SU(2)_s$ the magnon field transforms as 
$P(x)' = g P(x) g^\dagger$, while under the displacement $D_i$ by one lattice 
spacing in the $i$-direction (which changes the sign of the staggered 
magnetization) it transforms as $^{D_i}P(x) = \1 - P(x)$. Under spatial 
rotations $O$ and under reflections $R$ one obtains $^OP(x) = P(Ox)$ and
$^RP(x) = P(Rx)$ with $Ox = (- x_2,x_1,t)$ and $Rx = (x_1,- x_2,t)$. Finally,
under time-reversal $^TP(x) = \ ^{D_i}P(Tx)$ with $Tx = (x_1,x_2,- t)$.

In order to couple holes to the magnons, a nonlinear realization of the 
spontaneously broken $SU(2)_s$ symmetry has been constructed \cite{Kae05}. It
then manifests itself as a local symmetry in the unbroken $U(1)_s$ subgroup of
$SU(2)_s$. The local $U(1)_s$ transformations are constructed from the global 
transformation $g \in SU(2)_s$ as well as from the local magnon field $P(x)$.
First, one diagonalizes $P(x)$ by a unitary transformation $u(x) \in SU(2)_s$
\begin{equation}
u(x) P(x) u(x)^\dagger = \frac{1}{2}\left[\1 + \sigma_3\right], \ 
u_{11}(x) \geq 0.
\end{equation}
Under a global $SU(2)_s$ transformation $g$ the diagonalizing field $u(x)$
transforms as $u(x)' = h(x) u(x) g^\dagger$, which defines the nonlinear 
symmetry transformation  $h(x) = \exp(i \alpha(x) \sigma_3) \in U(1)_s$. Under 
the displacement symmetry $D_i$ one obtains $^{D_i}u(x) = \tau(x) u(x)$ with
\begin{equation}
\tau(x) = \left(\begin{array}{cc} 0 & - \exp(- i \varphi(x)) \\
\exp(i \varphi(x)) & 0 \end{array} \right).
\end{equation}
Next one constructs the anti-Hermitean composite field
\begin{equation}
v_\mu(x) = u(x) \p_\mu u(x)^\dagger =
i \left(\begin{array}{cc} v_\mu^3(x) &  v_\mu^+(x) \\ v_\mu^-(x) & - v_\mu^3(x)
\end{array}\right), 
\end{equation} 
which transforms under $SU(2)_s$ as
\begin{equation}
v_\mu(x)' = h(x) [v_\mu(x) + \p_\mu] h(x)^\dagger.
\end{equation}
The Abelian component $v_\mu^3(x)$ transforms like a $U(1)_s$ gauge field, 
while $v_\mu^\pm(x)$ represent ``charged'' vector fields. Under the 
displacement symmetry $D_i$ the composite vector field transforms as
$^{D_i}v_\mu(x) = \tau(x)[v_\mu(x) + \p_\mu] \tau(x)^\dagger$, while under 
time-reversal $^Tv_i(x) = \ ^{D_i}v_i(Tx)$ and 
$^Tv_t(x) = - \ ^{D_i}v_t(Tx)$.

In analogy to baryon chiral perturbation theory --- the effective theory for 
pions and nucleons \cite{Gas88,Jen91,Ber92,Bec99} --- a systematic low-energy 
effective theory was recently developed for magnons and holes \cite{Kae05}. 
The Hubbard model can be doped with both holes and electrons, and the $U(1)_Q$ 
fermion number symmetry is even extended to a non-Abelian $SU(2)_Q$ symmetry. 
For simplicity, here we consider underlying microscopic systems such as the 
$t$-$J$ model, for which the addition of electrons beyond half-filling is 
forbidden. Hence, we consider an effective theory with holes as the only charge
carriers. In \cite{Kae05} we have considered charge carriers located near
momenta $(0,0)$ and $(\frac{\pi}{a},\frac{\pi}{a})$ in the Brillouin zone 
(where $a$ is the lattice spacing). Here we consider hole pockets centered at 
$k^\alpha = (\frac{\pi}{2a},\frac{\pi}{2a})$ and 
$k^\beta = (\frac{\pi}{2a},- \frac{\pi}{2a})$, which have been observed
in ARPES measurements \cite{Wel95,LaR97,Kim98,Ron98} as well as in theoretical
calculations in the $t$-$J$ model \cite{Tru88,Els90,Bru00}. The hole fields 
$\psi^f_s(x)$ carry a ``flavor'' index $f = \alpha, \beta$ that characterizes
the corresponding hole pocket.  The index $s = \pm$ denotes spin parallel ($+$)
or antiparallel ($-$) to the local staggered magnetization. Under the various 
symmetry operations the hole fields transform as
\begin{eqnarray}
SU(2)_s:&&\psi^f_\pm(x)' = \exp(\pm i \alpha(x)) \psi^f_\pm(x),
\nonumber \\
U(1)_Q:&&^Q\psi^f_\pm(x) = \exp(i \omega) \psi^f_\pm(x),
\nonumber \\
D_i:&&^{D_i}\psi^f_\pm(x) = 
\mp \exp(i k^f_i a) \exp(\mp i \varphi(x)) \psi^f_\mp(x),
\nonumber \\
O:&&^O\psi^\alpha_\pm(x) = \mp \psi^\beta_\pm(Ox), \ 
^O\psi^\beta_\pm(x) = \psi^\alpha_\pm(Ox), \nonumber \\
R:&&^R\psi^\alpha_\pm(x) = \psi^\beta_\pm(Rx), \ 
^R\psi^\beta_\pm(x) = \psi^\alpha_\pm(Rx), \nonumber \\
T:&&^T\psi^f_\pm(x) = \mp \exp(\mp i \varphi(Tx)) \psi^{f\dagger}_\pm(Tx),
\nonumber \\
&&^T\psi^{f\dagger}_\pm(x) = \pm \exp(\pm i \varphi(Tx)) \psi^f_\pm(Tx).
\end{eqnarray}
Here $\omega$ determines the $U(1)_Q$ fermion number transformation.
Interestingly, in the effective theory the location $(k^f_1,k^f_2)$ of the hole
pockets in the Brillouin zone of the underlying crystal lattice manifests 
itself through charges $k^f_i a = \pm \frac{\pi}{2}$ of an internal Abelian 
symmetry $D_i$. Defining a $U(1)_s$ covariant derivative
\begin{equation}
D_\mu \psi^f_\pm(x) = \left[\p_\mu \pm i v_\mu^3(x)\right] \psi^f_\pm(x), 
\end{equation}
the leading terms in the effective Lagrangian are
\begin{eqnarray}
&&{\cal L}[\psi^{f\dagger}_s,\psi^f_s,P] = 
\rho_s \mbox{Tr}[\p_i P \p_i P + \frac{1}{c^2} \p_t P \p_t P] \nonumber \\
&&+ \sum_{f=\alpha,\beta; s = +,-}
[M \psi^{f\dagger}_s \psi^f_s + \psi^{f\dagger}_s D_t \psi^f_s + 
\frac{1}{2 M'} D_i \psi^{f\dagger}_s D_i \psi^f_s \nonumber \\
&&+ \sigma_f \frac{1}{2 M''} (D_1 \psi^{f\dagger}_s D_2 \psi^f_s +
D_2 \psi^{f\dagger}_s D_1 \psi^f_s) \nonumber \\
&&+ \Lambda (\psi^{f\dagger}_s v^s_1 \psi^f_{-s} 
+ \sigma_f \psi^{f\dagger}_s v^s_2 \psi^f_{-s}) +
N_1 \psi^{f\dagger}_s v^s_i v^{-s}_i \psi^f_s \nonumber \\
&&+ \sigma_f N_2 (\psi^{f\dagger}_s v^s_1 v^{-s}_2 \psi^f_s + 
\psi^{f\dagger}_s v^s_2 v^{-s}_1 \psi^f_s) 
\!+\!\frac{G_1}{2} \psi^{f\dagger}_s \psi^f_s \psi^{f\dagger}_{-s} \psi^f_{-s}]
\nonumber \\
&&+ \sum_{s = +,-} 
[G_2 \psi^{\alpha\dagger}_s \psi^\alpha_s \psi^{\beta\dagger}_s \psi^\beta_s +
G_3 \psi^{\alpha\dagger}_s \psi^\alpha_s 
\psi^{\beta\dagger}_{-s} \psi^\beta_{-s}].
\end{eqnarray}
Here $\rho_s$ is the spin stiffness, $c$ is the spinwave velocity, $M$, $M'$ 
and $M''$ are the rest mass and kinetic masses of a hole, $\Lambda$ is a
hole-one-magnon, and $N_1$ and $N_2$ are hole-two-magnon couplings, and $G_1$, 
$G_2$, and $G_3$ are 4-fermion contact interactions. All these low-energy
parameters take real values. The sign $\sigma_f$ is $+$ for 
$f = \alpha$ and $-$ for $f = \beta$. Interestingly, the above Lagrangian has 
an accidental $U(1)_F$ flavor symmetry that acts as
\begin{equation}
U(1)_F: \ ^F\psi^f_\pm(x) = \exp(\sigma_f i \eta) \psi^f_\pm(x).
\end{equation}
In addition, for $c \rightarrow \infty$ it also has an accidental Galilean 
symmetry. Both accidental symmetries are explicitly broken at higher orders of 
the derivative expansion.

Our treatment of the forces between two holes is analogous to the effective 
theory for light nuclei \cite{Wei90,Kap98,Epe98,Bed98} in which one-pion 
exchange dominates the long-range forces. We now calculate the one-magnon 
exchange potential. For this purpose, we expand in the magnon fluctuations 
$m_1(x)$, $m_2(x)$ around the ordered staggered magnetization, i.e.
\begin{eqnarray}
&&\vec e(x) = (\frac{m_1(x)}{\sqrt{\rho_s}},\frac{m_2(x)}{\sqrt{\rho_s}},1) + 
{\cal O}[m^2] \ \Rightarrow \ v_\mu^3(x) = {\cal O}[m^2], \nonumber \\
&&v_\mu^\pm(x) = \frac{1}{2 \sqrt{\rho_s}} \p_\mu [m_2(x) \pm i m_1(x)] + 
{\cal O}[m^2].
\end{eqnarray}
Since vertices with $v_\mu^3(x)$ (contained in $D_\mu$) involve at least two 
magnons, one-magnon exchange results from vertices with $v_\mu^\pm(x)$ only. 
As a consequence, two holes can exchange a single magnon only if they have
antiparallel spins ($+$ and $-$), which are both flipped in the magnon exchange
process. It is straightforward to evaluate the Feynman diagram describing 
one-magnon exchange shown in figure 1. 
\begin{figure}[tb]
\begin{center}
\vspace{-0.2cm}
\epsfig{file=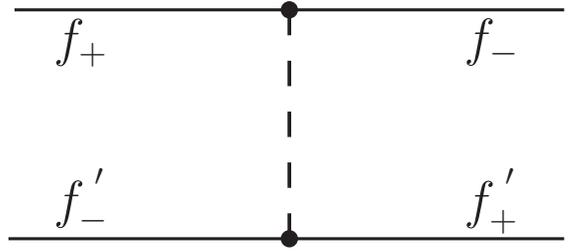,width=8cm}
\end{center}
\caption{\it Feynman diagram for one-magnon exchange between two holes with 
antiparallel spins undergoing a spin-flip.}
\vspace{-0.2cm}
\end{figure}

In coordinate space the resulting potentials for the various combinations of
flavors take the form
\begin{equation}
V^{ff}(\vec r) = \sigma_f \gamma \frac{\sin(2 \varphi)}{\vec r \, ^2}, \
V^{\alpha\beta}(\vec r) = \gamma \frac{\cos(2 \varphi)}{\vec r \, ^2},
\end{equation}
with $\gamma = \Lambda^2/(2 \pi \rho_s)$. Here $\vec r$ is the distance vector 
between the two holes and $\varphi$ is the angle between $\vec r$ and the
$x$-axis. It should be noted that the one-magnon exchange potential is 
instantaneous although magnons travel with the finite speed $c$. Retardation 
effects occur only at higher orders. A similar magnon exchange potential has 
been extracted directly from the $t$-$J$ model in \cite{Kuc93,Sus04}. In 
contrast to our method, that calculation is, however, affected by uncontrolled 
approximations.

Next we study the Schr\"odinger equation for the relative motion of two holes
with flavors $\alpha$ and $\beta$
\begin{equation}
\left(\begin{array}{cc} - \frac{1}{M'} \Delta & V^{\alpha\beta}(\vec r) \\
V^{\alpha\beta}(\vec r) &  - \frac{1}{M'} \Delta \end{array} \right)
\left(\begin{array}{c} \Psi_1(\vec r) \\ 
\Psi_2(\vec r) \end{array}\right) = E 
\left(\begin{array}{c} \Psi_1(\vec r) \\ 
\Psi_2(\vec r) \end{array}\right).
\end{equation}
The components $\Psi_1(\vec r)$ and $\Psi_2(\vec r)$ are probability
amplitudes for the spin-flavor combinations $\alpha_+\beta_-$ and 
$\alpha_-\beta_+$, respectively. The potential $V^{\alpha\beta}(\vec r)$ 
couples the two channels because magnon exchange is accompanied by a spin-flip.
The above Schr\"odinger equation does not yet account for the short-distance 
forces arising from 4-fermion contact interactions. Their effect will be 
incorporated later by a boundary condition on the wave function near the 
origin. Making the ansatz 
\begin{equation}
\Psi_1(\vec r) \pm \Psi_2(\vec r) = R(r) \chi_\pm(\varphi),
\end{equation}
for the angular part of the wave function one obtains
\begin{equation}
- \frac{d^2\chi_\pm(\varphi)}{d\varphi^2} \pm 
M' \gamma \cos(2 \varphi) \chi_\pm(\varphi) = - \lambda \chi_\pm(\varphi).
\end{equation}
The solutions of this Mathieu equation with the lowest eigenvalue $\lambda$
(given here only to the leading order $\gamma^2$) is
\begin{equation}
\chi_\pm(\varphi) = \frac{1}{\sqrt{\pi}}
\mbox{ce}_0(\varphi,\pm \frac{1}{2} M' \gamma), \ 
\lambda = \frac{1}{8} (M' \gamma)^2. 
\end{equation}
The periodic Mathieu function $\mbox{ce}_0(\varphi,\frac{1}{2} M' \gamma)$ 
\cite{Abr72} is illustrated for $M' \gamma = 2.5$ in figure 2.
\begin{figure}[tb]
\begin{center}
\vspace{-0.2cm}
\epsfig{file=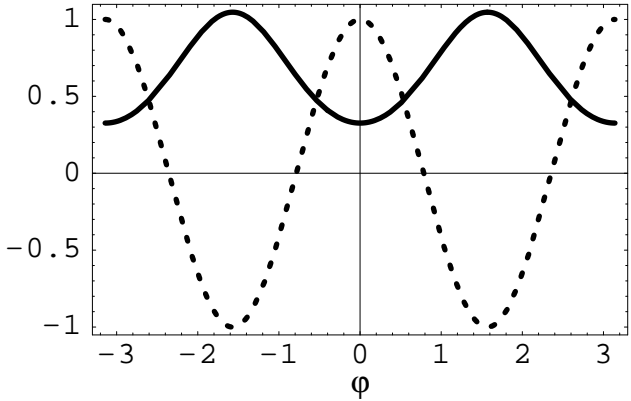,width=8.5cm}
\end{center}
\vspace{-0.2cm}
\caption{\it Angular wave function $\mbox{ce}_0(\varphi,\frac{1}{2} M' \gamma)$
(solid curve) and angle-dependence $\cos(2 \varphi)$ of the potential (dotted 
curve).}
\end{figure}

The radial Schr\"odinger equation takes the form
\begin{equation}
- \left[\frac{d^2R(r)}{dr^2} + 
\frac{1}{r} \frac{dR(r)}{dr}\right] - \frac{\lambda}{r^2} R(r) = M' E R(r).
\end{equation}
As it stands, the above equation is ill-defined because an attractive
$\frac{1}{r^2}$ potential is too singular at the origin. However, one should 
keep in mind that we have not yet incorporated the short-range contact 
interactions. A consistent description of the short-distance physics requires 
ultraviolet regularization and subsequent renormalization of the Schr\"odinger 
equation as discussed in \cite{Lep97}. Instead of proceeding systematically in 
this way (which will be the subject of a forthcoming publication), here we 
model the short-distance repulsion between two holes by a hard core of radius 
$r_0$, i.e.\ we require $R(r_0) = 0$. The radial Schr\"odinger equation for the
bound states is solved by Bessel functions
\begin{equation}
R(r) = A K_\nu(\sqrt{M' |E_n|} r), \ \nu = i \sqrt{\lambda}.
\end{equation}
The energy (determined from $K_\nu(\sqrt{M' |E_n|} r_0) = 0$) is given by
\begin{equation}
E_n \sim - (M' r_0^2)^{-1} \exp(- 2 \pi n/\sqrt{\lambda})
\end{equation}
for large $n$. While the highly excited states have exponentially small energy 
and exponentially large size, for sufficiently small $r_0$ or sufficiently 
large coupling $\Lambda$ the ground state is strongly bound. It should be noted
that the wave functions with angular part $\chi_+(\varphi)$ and 
$\chi_-(\varphi)$ have the same energy, i.e.\ the states are two-fold 
degenerate. Combining the two degenerate ground states to eigenstates of the
rotation $O$ one obtains the probability distribution illustrated in figure 3,
which resembles \,\, $d$-wave symmetry. It should, however, be noted that, due 
to the nontrivial rotation properties of flavor, the wave function is 
suppressed --- but not equal to zero --- along the lattice diagonals. 
\begin{figure}[tb]
\begin{center}
\vspace{-0.2cm}
\epsfig{file=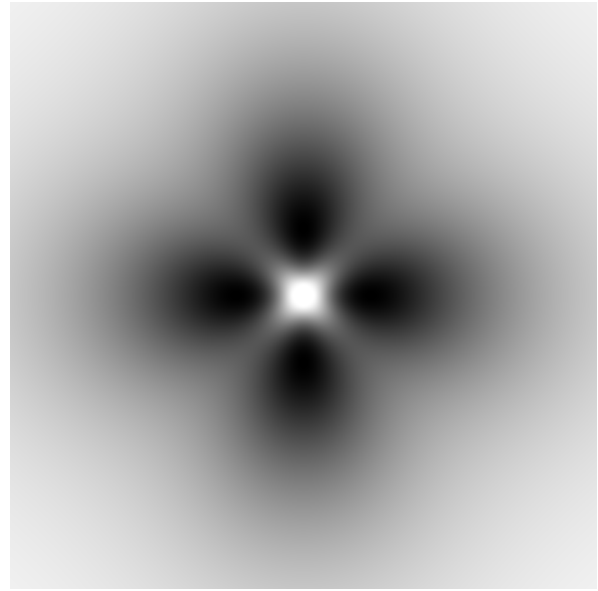,width=8.1cm}
\end{center}
\vspace{-0.2cm}
\caption{\it Probability distribution for the ground state of two holes with 
flavors $\alpha$ and $\beta$.}
\end{figure}
Pairs of holes with equal flavor can also form. The corresponding wave 
functions will be discussed elsewhere. Whether pairs of the same or of 
different flavors are more strongly bound depends on the values of the 
low-energy parameters. Here we have concentrated on $\alpha \beta$ pairs 
because they have important properties observed in the cuprates.

Since we have now established that two holes in an antiferromagnet can form a 
bound state by exchanging magnons, it is natural to ask if and how this may be
related to high-temperature superconductivity. Quantitatively these questions 
will be addressed elsewhere. Here we argue just qualitatively. 
If pairs of holes form bound states, at a sufficiently low temperature $T_c$ 
these pairs will condense, thus leading to superconductivity. Here we do not 
attempt to estimate $T_c$, because this involves a delicate interplay between 
long- and short-range interactions. Instead we concentrate on the universal 
aspects of the dynamics resulting from the long-range magnon-mediated forces 
only. First, one-magnon exchange only binds holes with antiparallel spins, and 
indeed the Cooper pairs in a high-temperature superconductor are spin singlets.
Second, the characteristic angular dependence $\cos(2 \varphi)$ of the 
one-magnon exchange potential leads to the peculiar 
$\mbox{ce}_0(\varphi,\frac{1}{2} M' \gamma)$ orbital structures of the hole 
pair wave function which yields the $d$-wave characteristics observed in the 
cuprates.

Besides basic principles of quantum field theory, such as locality and 
unitarity, the effective theory of magnons and holes relies only on a few 
experimentally well verified dynamical assumptions --- most important the 
spontaneous breaking of the $SU(2)_s$ spin symmetry down to $U(1)_s$ and the
location of hole pockets at $(\frac{\pi}{2a},\pm \frac{\pi}{2a})$. It is 
remarkable that the existence of bound states between holes in an 
antiferromagnet can be inferred from so little input. The effective theory
provides the detailed analytic form of the wave function for a pair of holes 
with different flavors $\alpha \beta$ which could perhaps be compared with 
experiments. While the corresponding probability distribution resembles 
$d$-wave characteristics, due to the nontrivial flavor structure the rotation 
symmetry $O$ is, nevertheless, realized in a more complicated way.

It is natural to ask if the effective theory can be applied to the 
high-temperature superconductors themselves. Since this theory relies 
on the spontaneous breakdown of the $SU(2)_s$ symmetry, and since 
high-temperature superconductivity arises only after antiferromagnetism has 
been destroyed, this may seem doubtful. However, while the perturbative 
treatment of the effective theory breaks down in the superconducting phase, the
effective theory itself does not, as long as spin fluctuations remain among
the relevant low-energy degrees of freedom. While it remains to be seen if 
nonperturbative investigations of the effective theory can shed light on the 
phenomenon of high-temperature superconductivity itself, it seems clear already
that the systematic low-energy effective field theory approach to the dynamics 
of charge carriers in antiferromagnets is promising.\\

We are indebted to J.\ Gasser for illuminating
discussions. We also thank the referees for constructive criticism of the
original manuscript. This work was supported in part by the Schweizerischer
Nationalfonds.

\end{document}